\documentclass[12pt,prd,aps,epsfig,floats,preprint,eqsecnum,superscriptaddress]{revtex4}
\usepackage{graphicx,amsmath}

\newcommand{\be}{\begin{equation}}
\newcommand{\ee}{\end{equation}}
\newcommand{\bea}{\begin{eqnarray}}
\newcommand{\eea}{\end{eqnarray}}
\newcommand{\beq}{\begin{equation}}
\newcommand{\eeq}{\end{equation}}
\newcommand{\beqa}{\begin{eqnarray}}
\newcommand{\eeqa}{\end{eqnarray}}
\newcommand{\no}{\nonumber}
\def\lsim{\mathrel{\rlap{\lower4pt\hbox{\hskip1pt$\sim$}}
     \raise1pt\hbox{$<$}}}         
\def\gsim{\mathrel{\rlap{\lower4pt\hbox{\hskip1pt$\sim$}}
     \raise1pt\hbox{$>$}}}         
\newcommand{\ddbar}{D^0-\overline{D}{}^0}

\begin{document}


\vspace*{.0cm}

\title{Lessons from Recent Measurements of $\ddbar$ Mixing}

\author{Oram Gedalia} \email{oram.gedalia@weizmann.ac.il}
\affiliation{Department of Particle Physics,
   Weizmann Institute of Science, Rehovot 76100, Israel}
\author{Yuval Grossman}\email{yg73@cornell.edu}
\affiliation{Institute for High Energy Phenomenology, Newman
  Laboratory of Elementary Particle Physics, Cornell University,
  Ithaca, NY 14853, USA}
\author{Yosef Nir\footnote{The Amos de-Shalit chair of theoretical
     physics}}\email{yosef.nir,@weizmann.ac.il}
\author{Gilad Perez}\email{gilad.perez@weizmann.ac.il}
\affiliation{Department of Particle Physics,
   Weizmann Institute of Science, Rehovot 76100, Israel}

\vspace*{1cm}
\begin{abstract}
An impressive progress in measurements of the $\ddbar$ mixing
parameters has been made in recent years. We explore the implications
of these measurements to models of new physics, especially in view of
recent upper bounds on the amount of CP violation. We update the
constraints on non-renormalizable four-quark operators. We show that
the experiments are close to probing minimally flavor violating models
with large $\tan\beta$. The data challenge models with a scale of
order TeV where the flavor violation in the down sector is suppressed
by alignment and, in particular, certain classes of supersymmetric
models and of warped extra dimension models.
\end{abstract}
\maketitle

\section{Introduction}
The neutral $D$-meson system is the only one among the four neutral
meson systems ($K,D,B,B_s$) that is made of up-type quarks. But this
is not the only unique aspect of this system:
\begin{enumerate}
\item It is the only system where long distance contributions to
the mixing are orders of magnitude above the (Standard Model) short
distance ones.
\item It is the only system where the Standard Model (SM) contribution
to the CP violation in the mixing amplitude is expected to be below
the permil level.
\end{enumerate}
The first point means that it is extremely difficult to theoretically
predict the width- and (more significant for our purposes)
mass-splitting. The second point implies that, in spite of this
inherent uncertainty, $\ddbar$ mixing can unambiguously signal new
physics in case that it is found to exhibit CP violation.
Alternatively, if no CP violation is observed, it will provide useful
constraints on new physics .

It is the purpose of this paper to obtain constraints on new physics
from the new experimental data.  In particular, the fact that there is
now, for the first time, evidence that CP violation in $\ddbar$ mixing
is small, can be cleanly interpreted in various frameworks of new
physics.

The plan of this paper is as follows. In Section \ref{sec:exp} we
define the experimental parameters and present their allowed ranges.
In Section \ref{sec:the} we define the theoretical parameters and
derive their allowed ranges. In Section \ref{sec:ind} we apply the
bounds to a generic effective theory, namely to $\Delta C=2$ four
quark operators, while in Section \ref{sec:mfv} we focus on minimally
flavor violating models. Sections \ref{sec:sus} and \ref{sec:rs} deal
with, respectively, supersymmetric models of alignment and models of
warped extra dimensions. Our conclusions are summarized in Section
\ref{sec:con}.

\section{The experimental parameters}
\label{sec:exp}
We start be reviewing the formalism of charm mixing (see, for example,
\cite{Blaylock:1995ay,Golowich:2007ka,Bianco:2003vb,Bigi:2009df}). The
two neutral $D$-meson mass eigenstates, $|D_1\rangle$ of mass $m_1$
and width $\Gamma_1$ and $|D_2\rangle$ of mass $m_2$ and width
$\Gamma_2$, are linear combinations of the interaction eigenstates
$|D^0\rangle$ (with quark content $c\bar u$) and
$|\overline{D^0}\rangle$ (with quark content $\bar cu$):
\beq
|D_1\rangle=p|D^0\rangle+q|\overline{D^0}\rangle,\qquad
|D_2\rangle=p|D^0\rangle-q|\overline{D^0}\rangle,
\eeq
where under CP transformation $D^0$ and $\overline{D^0}$ are interchanged.
The average and the difference in mass and width are given by
\beqa
m\equiv\frac{m_1+m_2}{2},&\ \ \ \
&\Gamma\equiv\frac{\Gamma_1+\Gamma_2}{2},\no\\
x\equiv\frac{m_2-m_1}{\Gamma},&\ \ \ \
&y\equiv\frac{\Gamma_2-\Gamma_1}{2\Gamma}.
\eeqa
The decay amplitudes into a final state $f$ are defined as follows:
\beq
A_f=\langle f|{\cal H}|D^0\rangle,\qquad
\overline{A}_f=\langle f|{\cal H}|\overline{D^0}\rangle.
\eeq
We define a complex dimensionless parameter $\lambda_f$:
\beq
\lambda_f=\frac qp \frac{\overline{A}_f}{A_f}.
\eeq
The time-dependent decay rates of interest are those of the
doubly-Cabibbo-suppressed decay into a flavor-specific final
state.
\beqa\label{eq:dcs}
\Gamma[D^0(t)&\to&K^+\pi^-]=e^{-\Gamma t}|\overline{A}_{K^+\pi^-}|^2
|q/p|^2\no\\
&\times&\left\{|\lambda^{-1}_{K^+\pi^-}|^2
+[{\cal R}e(\lambda^{-1}_{K^+\pi^-})y
+{\cal I}m(\lambda^{-1}_{K^+\pi^-})x]\Gamma t
+\frac14(y^2+x^2)(\Gamma t)^2\right\},\no\\
\Gamma[\overline{D^0}(t)&\to&K^-\pi^+]=e^{-\Gamma t}|A_{K^-\pi^+}|^2
|p/q|^2\no\\
&\times&\left\{|\lambda_{K^-\pi^+}|^2
+[{\cal R}e(\lambda_{K^-\pi^+})y
+{\cal I}m(\lambda_{K^-\pi^+})x]\Gamma t
+\frac14(y^2+x^2)(\Gamma t)^2\right\},
\eeqa
and singly-Cabibbo-suppressed decay into a CP eigenstate:
\beqa\label{eq:scs}
\Gamma[D^0(t)\to K^+K^-]&=&e^{-\Gamma t}|A_{K^+K^-}|^2
\left\{1+[{\cal R}e(\lambda_{K^+K^-})y
-{\cal I}m(\lambda_{K^+K^-})x]\Gamma t\right\},\no\\
\Gamma[\overline{D^0}(t)\to K^+K-]&=&e^{-\Gamma t}
|\overline{A}_{K^+K^-}|^2
\left\{1+[{\cal R}e(\lambda^{-1}_{K^+K^-})y
-{\cal I}m(\lambda^{-1}_{K^+K^-})x]\Gamma t\right\}.
\eeqa
The expressions above are valid only in the limit $x,y \ll 1$, which
is the case for the $D$ system.

 The effects of indirect CP violation can be parameterized
in the following way \cite{Bergmann:2000id}:
\beq\label{defphi}
\lambda^{-1}_{K^+\pi^-}= r_d\left|{\frac{p}{q}}\right|
e^{-i(\delta_{K\pi}+\phi)},\qquad \lambda_{K^-\pi^+}=
r_d\left|{\frac{q}{p}}\right|e^{-i(\delta_{K\pi}-\phi)},\qquad
\lambda_{K^+K^-}=-\left|{\frac{q}{p}}\right|e^{i\phi},
\eeq
where $r_d$ is a real and positive dimensionless parameter,
$\delta_f$ is a strong (CP conserving) mode-dependent phase, and
$\phi$ is a weak (CP violating) universal phase. Similar expressions can
be written to decays into any final state. The appearance of a single
weak phase that is common to all final states is related to the
absence of direct CP violation, while the absence of a strong phase in
$\lambda_{K^+K^-}$ is related to the fact that the final state is a CP
eigenstate.

In our analysis we assume that effects of direct CP violation are negligibly small
even in the presence of new physics (NP).
The question of NP contributions to direct CP violation in the doubly Cabibbo suppressed
decays was investigated in detail in \cite{Bergmann:1999pm,D'Ambrosio:2001wg}
and shown to be indeed generically small. In some special cases it
could reach order 30\%.
The singly Cabibbo suppressed decays case was studied in~\cite{Grossman:2006jg}. Typically direct CP violation
 is suppressed, but in special
models (or corners of parameter space) it could be non-negligible.
Experimental constraints on direct CP violation in charm decays were analyzed by the
heavy flavor averaging group (HFAG)~\cite{Barberio:2008fa} and found to be
of order one percent. Furthermore, the effect of including direct CP violation on the NP contributions was recently considered in~\cite{Kagan:2009gb} and shown to be subdominant.

The experimental measurements of the various relevant $D$-decay
rates can be used to determine the values of the four
parameters that are related to $D^0-\overline{D^0}$ mixing:
$x,y,|q/p|$ and $\phi$. Impressive progress in relevant
measurements has been recently achieved in the BaBar and Belle
experiments. The information comes from a variety of final
states of neutral $D$-meson decays: $K^+K^-$, $\pi^+\pi^-$,
$K\pi^+\pi^-$, $K\ell\nu$, $K^-\pi^+$ and $K^+\pi^-$. HFAG has fitted the data, and obtained
the following one sigma ranges~\cite{Barberio:2008fa}:
\beqa\label{hfagexp}
x&=&(1.00\pm0.25)\times10^{-2},\no\\
y&=&(0.77\pm0.18)\times10^{-2},\no\\
1-|q/p|&=&+0.06\pm0.14,\no\\
\phi&=&-0.05\pm0.09,
\eeqa
where $\phi$ is given in radians.
These results imply the following:
\begin{enumerate}
\item The width-splitting and mass-splitting are at a level
close to one percent.
\item CP violation is small.
\end{enumerate}
We would now like to translate these statements, made for the parameters
that are used to describe the experimental results, to parameters that
represent the theory input.

\section{The Theoretical Parameters}
\label{sec:the}
The $\overline{D^0}-D^0$ transition amplitudes are defined as follows:
\beq
\langle D^0|{\cal H}|\overline{D^0}\rangle=M_{12}-\frac i2\Gamma_{12},\qquad
\langle\overline{D^0}|{\cal H}|D^0\rangle=M_{12}^*-\frac i2\Gamma_{12}^*.
\eeq
The overall phase of the mixing amplitude is not a physical
quantity. It can be changed by the choice of phase convention for
the up and charm quarks. The relative phase between $M_{12}$ and
$\Gamma_{12}$ is, however, phase convention independent and has
therefore physics consequences.  The three physical quantities
related to the mixing can be defined as
\beq\label{thepar}
y_{12}\equiv|\Gamma_{12}|/\Gamma,\qquad
x_{12}\equiv2|M_{12}|/\Gamma,\qquad
\phi_{12}\equiv\arg(M_{12}/\Gamma_{12}).
\eeq
Note that various papers use different sign conventions for $\phi_{12}$.
In the absence of direct CP violation, the following two conditions
are met:
\beq
{\cal I}m(\Gamma_{12}^*\overline{A}_f/A_f)=0,\qquad
|\overline{A}_f/A_f|=1.
\eeq

In the following we assume that the tree level decay amplitudes in the
processes (\ref{eq:dcs}) and (\ref{eq:scs}) are given by the SM. This
implies that there is no direct CP violation and that $y_{12}$, which
is generated by decay into final states that are common to $D^0$ and
$\overline{D^0}$ decays, is described by SM physics (see,
however~\cite{Golowich:2006gq}).  In that case the relations between
the experimental parameters and the theoretical ones are given in Ref.
\cite{Grossman:2009mn}. Given a new physics model, one can calculate
$x_{12}$ and $\phi_{12}$ in terms of the model parameters. We are thus
particularly interested in using experimental data to constrain
$x_{12}$ and $\phi_{12}$ and subsequently the new physics model
parameters. Actually, the parameters that are most convenient for the
analysis of NP effects are $x_{12}$ and
$x_{12}\sin\phi_{12}$ which are related directly to, respectively, the
absolute value and the imaginary part of the NP operators.

We can express the theoretical parameters in terms of $x,y$ and
$\phi$,
\beqa\label{the:phi}
x_{12}^2&=&\frac{x^4\cos^2\phi+y^4\sin^2\phi}
{x^2\cos^2\phi-y^2\sin^2\phi},\no\\
\sin^2\phi_{12}&=&\frac{(x^2+y^2)^2\cos^2\phi\sin^2\phi}
{x^4\cos^2\phi+y^4\sin^2\phi},
\eeqa
or $x,y$ and $|q/p|$,
\beqa\label{xphicp}
x_{12}^2&=&x^2\frac{(1+|q/p|^2)^2}{4|q/p|^2}+
y^2\frac{(1-|q/p|^2)^2}{4|q/p|^2},\no\\
\sin^2\phi_{12}&=&\frac{(x^2+y^2)^2(1-|q/p|^4)^2}
{16x^2y^2|q/p|^4+(x^2+y^2)^2(1-|q/p|^4)^2}.
\eeqa
The fact that the four experimental observables ($x,y,\phi,|q/p|$) can
be expressed in terms of three theoretical input parameters
($x_{12},y_{12},\phi_{12}$) means that a model-independet relation
between the experimental observables is predicted
\cite{Grossman:2009mn}.

Using the relations between the experimental and theoretical
parameters, the experimental data bound the theoretical
parameters. Since the relations are complicated, we need the full
correlations between the experimental measurements to perform such a
task. We do not aim to do it here; Instead, we assume no
correlation and use the $1\sigma$ bounds. Once a full treatment is
done, our results can be straightforwardly re-scaled to the new
bounds.

There are two common parameterizations that are used
to present constraints on new physics from neutral meson mixing.
These are the $(r_D,\theta_D)$ and the $(h_D,\sigma_D)$
parameterizations defined by
\beq
r_D^2=(x_{12}/x_{12}^{\rm SM}),\qquad
2\theta_D=\phi_{12}-\phi_{12}^{\rm SM}; \qquad
h_D=(x_{12}^{\rm NP}/x_{12}^{\rm SM}), \qquad
2\sigma_D=\phi_{12}^{\rm NP}-\phi_{12}^{\rm SM}.
\eeq
We use the superscript ``SM'' (``NP'') to denote the Standard Model
(New Physics) contributions. These parameterizations are, however,
inappropriate for the discussion of $D^0-\overline{D^0}$ mixing,
because $x_{12}^{\rm SM}$ is poorly known. It is more useful to use,
instead, the ratio $x_{12}^{\rm NP}/x_{12}$ and the phase
$\phi_{12}^{\rm NP}$. In terms of the standard parameterizations,
$x_{12}^{\rm NP}/x_{12}=h_D/r_D^2$ and $\phi_{12}^{\rm NP}=2\sigma_D$
(where, for the latter equation, we use $\phi_{12}^{\rm SM}=0$).

The most plausible mechanism that has been identified as a possible
source of $y_{12}\sim0.01$ -- SU(3)-breaking in phase-space factors
\cite{Falk:2001hx} -- predicts that $x_{12}^{\rm SM}\lsim y_{12}$
\cite{Falk:2004wg}. It is therefore very likely that $x_{12}^{\rm
  SM}\lsim x_{12}$. Assuming that there are no accidental strong
cancellations between the standard model and the new physics
contributions to $M_{12}$, we can use the data to bound the NP
parameters.  The experimental data, Eq. (\ref{hfagexp}), then give
\beq\label{connp}
x_{12}^{\rm NP}\lsim x_{12}^{\rm exp}\sim 0.012,\qquad
x_{12}^{\rm NP}\sin\phi_{12}^{\rm NP}\lsim x_{12}^{\rm
exp}\sin\phi_{12}^{\rm exp} \sim 0.0022,
\eeq
where $x_{12}^{\rm exp}$ and $\sin\phi_{12}^{\rm exp}$ denote the
upper bounds on these theoretical parameters extracted from the
experimental data (at $1\sigma$, as explained above).  We plot
the constraints (\ref{connp}) in the $x_{12}^{\rm NP}/x_{12}-
\phi_{12}^{\rm NP}$ plane in Fig. 1 (the allowed region is
shown in grey).

\begin{figure}
\includegraphics[scale=1.0]{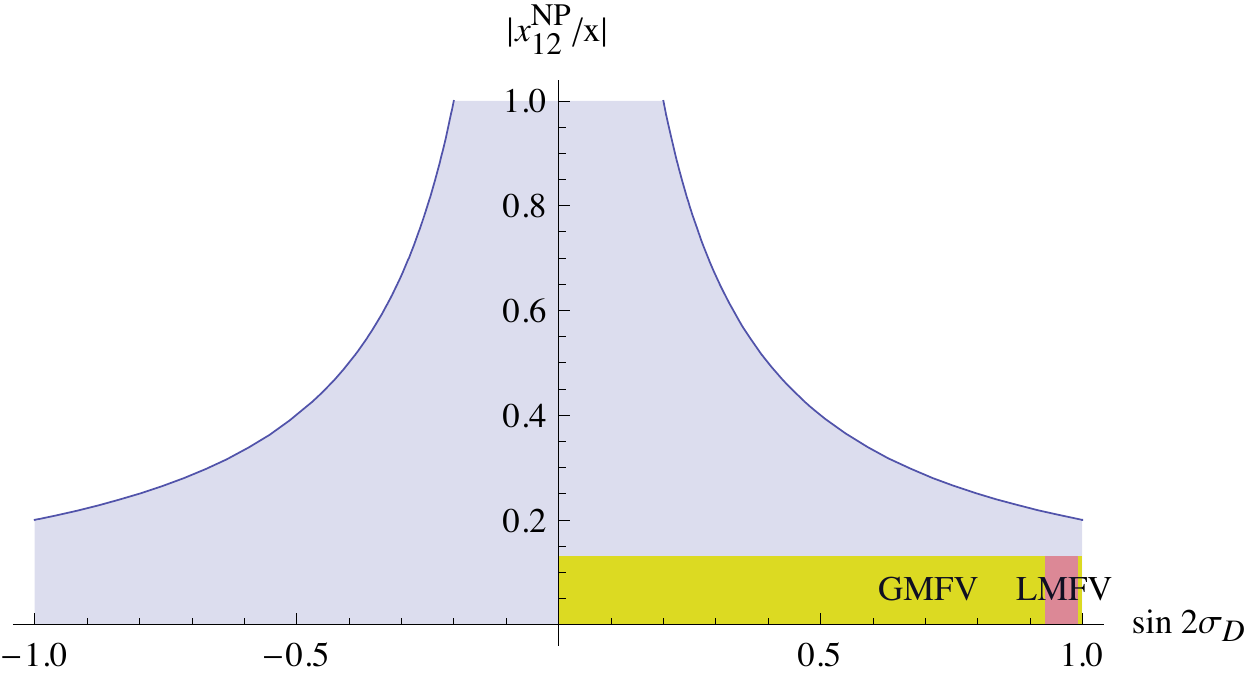}
\caption{\label{fig:connp} The allowed region, shown in grey, in the
$x_{12}^{\rm NP}/x_{12}-\sin\phi_{12}^{\rm NP}$ plane.
The pink and yellow regions correspond to the ranges predicted by,
respectively, the linear MFV and general MFV classes of models (see
section~\ref{sec:mfv} for details).}
\end{figure}

We now implement these constraints in various theoretical extensions
of the standard model.

\section{Model independent constraints}
\label{sec:ind}
The most general effective Hamiltonian for $\Delta C=2$ processes
(see~\cite{Golowich:2009ii} for a recent discussion on the $\Delta
C=1$ case) from new physics at a high scale $\Lambda_{\rm NP}\gg m_W$
can be written as follows:
\beq\label{nonren}
{\cal H}_{\rm eff}^{\Delta C=2}=\frac{1}{\Lambda_{\rm NP}^2}
\left(\sum_{i=1}^5 z_i Q_i^{cu}
  +\sum_{i=1}^3\tilde z_i \tilde Q_i^{cu}\right),
\eeq
where
\beqa
Q_1^{cu}&=&\bar u^\alpha_L\gamma_\mu c^\alpha_L\bar
u^\beta_L\gamma^\mu c^\beta_L,\no\\
Q_2^{cu}&=&\bar u^\alpha_R c^\alpha_L\bar
u^\beta_R c^\beta_L,\no\\
Q_3^{cu}&=&\bar u^\alpha_R c^\beta_L\bar
u^\beta_R c^\alpha_L,\no\\
Q_4^{cu}&=&\bar u^\alpha_R c^\alpha_L\bar
u^\beta_L c^\beta_R,\no\\
Q_5^{cu}&=&\bar u^\alpha_R c^\beta_L\bar
u^\beta_L c^\alpha_R,
\eeqa
and $\alpha,\beta$ are color indices. The operators $\tilde
Q^{cu}_{1,2,3}$ are obtained from the $Q^{cu}_{1,2,3}$ by the
exchange $L\leftrightarrow R$. In the following we only discuss
the operators $Q_i$ as the results for $Q_{1,2,3}$ apply to
$\tilde Q_{1,2,3}$ as well.

We take into account the running and mixing of the operators between
the scale of new physics and the $m_D$ scale. This is performed using
the formula
\beq
\langle \overline{D^0}|\mathcal{H}_{\mathrm{eff}}^{\Delta
F=2}|D^0\rangle_i=\sum_{j=1}^5 \sum_{r=1}^5 \left(
b_j^{(r,i)}+\eta c_j^{(r,i)} \right) \eta^{a_j} C_i({\Lambda})
\langle \overline{D^0}|Q_r^{cu}|D^0\rangle
\eeq
and the relevant inputs given in~\cite{Bona:2007vi,Ciuchini:2007cw}.

In this way, we obtain the following constraints from
$x_{12}^{\rm NP}\lsim0.012$:
\beqa\label{conzi}
|z_1|&\lsim&5.7\times10^{-7}
\left(\frac{\Lambda_{\rm NP}}{1\ TeV}\right)^2,\no\\
|z_2|&\lsim&1.6\times10^{-7}
\left(\frac{\Lambda_{\rm NP}}{1\ TeV}\right)^2,\no\\
|z_3|&\lsim&5.8\times10^{-7}
\left(\frac{\Lambda_{\rm NP}}{1\ TeV}\right)^2,\no\\
|z_4|&\lsim&5.6\times10^{-8}
\left(\frac{\Lambda_{\rm NP}}{1\ TeV}\right)^2,\no\\
|z_5|&\lsim&1.6\times10^{-7} \left(\frac{\Lambda_{\rm NP}}{1\
TeV}\right)^2.
\eeqa
We further obtain the following constraints from $x_{12}^{\rm
NP}\sin\phi_{12}^{\rm NP}\lsim0.0022$:
\beqa\label{conizi}
{\cal I}m(z_1)&\lsim&1.1\times10^{-7}
\left(\frac{\Lambda_{\rm NP}}{1\ TeV}\right)^2,\no\\
{\cal I}m(z_2)&\lsim&2.9\times10^{-8}
\left(\frac{\Lambda_{\rm NP}}{1\ TeV}\right)^2,\no\\
{\cal I}m(z_3)&\lsim&1.1\times10^{-7}
\left(\frac{\Lambda_{\rm NP}}{1\ TeV}\right)^2,\no\\
{\cal I}m(z_4)&\lsim&1.1\times10^{-8}
\left(\frac{\Lambda_{\rm NP}}{1\ TeV}\right)^2,\no\\
{\cal I}m(z_5)&\lsim&3.0\times10^{-8} \left(\frac{\Lambda_{\rm
NP}}{1\ TeV}\right)^2.
\eeqa
We learn the following points (the strongest constraints
correspond to maximal CP violating phases):
\begin{enumerate}
\item Generic new physics that contributes to the operators
    (\ref{nonren}) at tree level with couplings of ${\cal
    O}(1)$ must lie at a very high scale, $\Lambda_{\rm
    NP}\gsim (4-10)\times10^3\ TeV$.
\item Generic new physics that contributes to the operators
    (\ref{nonren}) at the loop level with effective
    coupling of ${\cal O}(\alpha_2^2)$ (similar to the SM)
    must lie at a high scale, $\Lambda_{\rm NP}\gsim
    (1-3)\times10^2\ TeV$.
\item New physics at or below the TeV scale must have a
    highly suppressed coupling, {\it e.g.},
    $z_1\lsim(1-6)\times10^{-7}$ and
    $z_4\lsim(1-6)\times10^{-8}$.
If the new physics contribution is loop-suppressed by
${\cal O}(\alpha_2^2)$, then the flavor suppression has to be
as strong as $10^{-4}-10^{-5}$.
\end{enumerate}

\section{Minimal flavor violation}
\label{sec:mfv} In models of minimal flavor violation
(MFV)~\cite{MFV,GMFV}, the Yukawa matrices are taken to be
spurions,
\beq
Y_u(3,\bar3,1),\ \ \ \ Y_d(3,1,\bar3),
\eeq
under the flavor symmetry
\beq
G_{\rm flavor}^q=SU(3)_Q\times SU(3)_U\times SU(3)_D.
\eeq
We are interested in contributions of MFV new physics to the
operators (\ref{nonren}). These operators generate transitions
between the first two generations of up mass eigenstates. The
relevant basis is then the up mass basis, where $Y_u$ is
diagonal. Since, however, $SU(3)_D$ is unbroken by the
operators (\ref{nonren}), all couplings $z_i$ must involve
powers of $Y_d Y_d^\dagger$. Note, however, that all operators
except $Q_1^{cu}$ break an $SU(2)_U$ subgroup of $SU(3)_U$, and
therefore the corresponding $z_i$ and $\tilde z_i$ are further
suppressed by, at least, $m_c^2/m_t^2$~\cite{GMFV}. We conclude
that, if the new physics operators arise with a similar
suppression scale in MFV models, the leading contribution is to
the $Q_1^{cu}$ operator:
\beqa\label{zmfv}
z_1\propto
[y_s^2(V_{cs}^*V_{us})+(1+r_{\rm GMFV})\times y_b^2(V_{cb}^*V_{ub})]^2
&\sim&\left\{
\begin{array}{cc} 1\times10^{-15}\zeta_1 & {\rm 1HDM} \cr
3\times10^{-8}\zeta_1 & {\rm 2HDM,}\ \tan\beta\sim m_t/m_b \cr
\end{array} \right.,
\eeqa
where $r_{\rm GMFV}$ is relevant to general MFV (GMFV) models,
in which the contributions from higher powers of the bottom
Yukawa coupling are important and need to be resummed. In such
a case, the simple relation between the contribution from the
strange and bottom quarks does not apply~\cite{GMFV}. In cases
where either $\tan\beta$ is low or only the leading term in the
MFV expansion is important, denoted as linear MFV (LMFV),
$r_{\rm GMFV}=0$; Otherwise it is expected to be an order one
number. We thus have
\beqa
\zeta_1&=& e^{2i\gamma} + 2r_{sb} e^{i\gamma}+r_{sb}^2\sim1.7i+r_{\rm
  GMFV}\left[2.4 i-1-0.7 \,r_{\rm GMFV}\left(1+i\right)\right],\no\\
r_{sb}&=&\frac{y_s^2}{y_b^2}\left|\frac{V_{us}V_{cs}}{V_{ub}V_{cb}}\right|\sim0.5,
\eeqa
where $\gamma$ is the relevant phase of the unitarity triangle.
We put here $y_d^2/y_s^2=0$ and use
$|V_{cb}V_{ub}|=1.8\times10^{-4}$ for the
Cabibbo-Kobayashi-Maskawa (CKM) matrix elements,
$\gamma\approx67^o$ and, within the SM, $y_b(1\ TeV)=0.014$
\cite{Xing:2007fb}. 1HDM stands for single Higgs doublet
models, such as the SM, while 2HDM stands for two Higgs doublet
models, such as the minimal supersymmetric standard model. For
the latter we assume that, for $y_b\approx y_t$, there is no
suppression from the bottom Yukawa coupling, namely
$y_b^4/\Lambda_{\rm NP}^2\approx 1/(1\ TeV)^2$. As concerns the
phase of the MFV contribution, there is a clear prediction for
the LMFV case, independent of whether we have a 1HDM or 2HDM
and of $\tan\beta$:
\beqa
\left.\frac{{\cal R}e(z_1)}{{\cal I}m(z_1)}\right|_{\rm
  LMFV}&=&\left.\frac{{\cal R}e(\zeta_1)}{{\cal
    I}m(\zeta_1)}\right|_{\rm
LMFV}=\frac{\cos2\gamma+2r_{sb}\cos\gamma+r_{sb}^2}
{\sin 2\gamma+2r_{sb}\sin\gamma+r_{sb}^2}\approx0,
\eeqa
where we use $y_b/y_s\sim53$ \cite{Xing:2007fb},
$|(V_{ub}V_{cb})/(V_{us}V_{cs})|\sim8.2\times10^{-4}$, and
$\sin2\gamma\sim0.7$.

We learn the following points:
\begin{enumerate}
\item MFV models with two Higgs doublets can contribute to
  $D^0-\overline{D^0}$ mixing up to ${\cal O}(0.1)$ for very large
  $\tan\beta$ (see also~\cite{GMFV}).
\item Single Higgs doublet models, and two Higgs doublet models with
  small $\tan\beta$, contribute at ${\cal O}(10^{-7})$.
\item The CP violating part of these contributions is not suppressed
  compared to the CP conserving part.
\end{enumerate}
Note that the CP violating part of these contributions is the
only part that can provide a convincing signal. It can give an
${\cal O}(0.1)$ effect, provided that two conditions are
fulfilled: First, $\tan\beta\sim m_t/m_b$ is required. Second,
the new physics contribution is either tree level or, if it is
loop level, logarithmically enhanced. The latter applies, for
example, to supersymmetry with gauge mediation with a
relatively high mediation scale (that is, not far below the
grand unification scale). For $\tan\beta\sim1$ (or single Higgs
doublet) models, the new physics effects are unobeservably
small. In Fig. \ref{fig:connp} we show in pink (yellow) the
range predicted by the LMFV (GMFV) class of models. The GMFV
yellow band is obtained by scanning the range $r_{\rm
GMFV}\in(-1,+1)$ (but keeping the magnitude of $z_1$ fixed for
simplicity).

\section{Supersymmetry}
\label{sec:sus}
$D^0-\overline{D^0}$ mixing constitutes a crucial test of
supersymmetric models with alignment
\cite{Nir:1993mx,Leurer:1993gy,Ciuchini:2007cw,Nir:2007ac}.  The
combination of $D^0-\overline{D^0}$ mixing and $K^0-\overline{K^0}$
mixing provides (within the supersymmetric framework) a model
independent constraint on the first two generations of squark doublets
\cite{Nir:2007xn,Feng:2007ke,Blum:2009sk}. In particular, it implies
that some level of degeneracy between their masses is required.

We ignore contributions involving squarks of the third
generations and $\tilde u_{L}-\tilde u_{R}$ mixing. We
parameterize the flavor suppression by the factors
\beqa
\delta_{LL}&=&\frac{\Delta\tilde m^2_{Q_2Q_1}}{\tilde
  m^2_Q}(K_{21}^{u_L}K_{11}^{u_L*}),\no\\
\delta_{RR}&=&\frac{\Delta\tilde m^2_{U_2U_1}}{\tilde
  m^2_U}(K_{21}^{u_R}K_{11}^{u_R*}),\no\\
\langle\delta\rangle^2&=&\delta_{LL}\delta_{RR}.
\eeqa
where $\tilde m_Q$ ($\tilde m_U$) is the average mass of the
first two squark up-doublet (up-singlet) generations,
$\Delta\tilde m^2_{Q_2Q_1}$ ($\Delta\tilde m^2_{U_2U_1}$) is
the mass-squared difference between them and $K^{u_L}$
($K^{u_R}$) is the mixing matrix in the gluino couplings to
up-quark doublet (singlet) and squark doublet (up-singlet)
pairs. The contributions from the first two up-type squark
generations to the various terms of Eq. (\ref{nonren}) can be
written as follows:
\beq
\begin{split}
\Lambda_{\rm NP}&=\tilde m,\\
z_1&=-\frac{\alpha_s^2}{216}g_1(m_{\tilde g}^2/\tilde m_Q^2)
\delta_{LL}^2,\\
\tilde z_1&=-\frac{\alpha_s^2}{216}g_1(m_{\tilde g}^2/\tilde
m_U^2) \delta_{RR}^2,\\
z_4&=-\frac{\alpha_s^2}{216}g_4(m_{\tilde g}^2/\tilde m^2)
\langle\delta\rangle^2,\\
z_5&=-\frac{\alpha_s^2}{216}g_5(m_{\tilde g}^2/\tilde m^2)
\langle\delta\rangle^2.
\end{split}
\eeq
Here, $\tilde m$ is the average squark mass, $m_{\tilde g}$ is
the gluino mass and $g_i(x)$ are known kinematic
functions~\cite{Ciuchini:1998ix} (for simplicity, we neglect
the $LR$ contributions),
\beq
\begin{split}
g_1(x)&=24x f_6(x)+66 \tilde{f}_6(x) \, ,\\
g_4(x)&=504x f_6(x)-72 \tilde{f}_6(x) \, ,\\
g_5(x)&=24x f_6(x) +120 \tilde{f}_6(x) \, ,
\end{split}
\eeq
with
\beq
\begin{split}
f_6(x)&=\frac{6(1+3x)\log{x}+x^3-9x^2-9x+17}{6(x-1)^5} \, ,
\\ \tilde{f}_6(x)&=\frac{6x(1+x)\log{x}-x^3-9x^2+9x+1}{
3(x-1)^5} \, .
\end{split}
\eeq

We use as reference point $m_{\tilde g}=\tilde m$, for which
\beq
g_1(1)=-1,\ \ \ g_4(1)=138/5,\ \ \ g_5(1)=-14/5.
\eeq
We obtain:
\beqa
z_1&\sim&3.7\times10^{-5}\delta_{LL}^2,\no\\
\tilde z_1&\sim&3.7\times10^{-5}\delta_{RR}^2,\no\\
z_4&\sim&1.0\times10^{-3}\langle\delta\rangle^2,\no\\
z_5&\sim&1.0\times10^{-4}\langle\delta\rangle^2.
\eeqa
Taking $\tilde m\lsim1\ TeV$, we find the following constrains from
Eqs. (\ref{conzi}) and  (\ref{conizi}):
\beqa\label{upabsd}
|\delta_{LL}|&\lsim&0.13,\no\\
|\delta_{RR}|&\lsim&0.13,\no\\
|\langle\delta\rangle|&\lsim&0.008,
\eeqa
\beqa\label{upimad}
\left[{\cal I}m(\delta_{LL})^2\right]^{1/2}&\lsim&0.05,\no\\
\left[{\cal I}m(\delta_{RR})^2\right]^{1/2}&\lsim&0.05,\no\\
\left[{\cal I}m\langle\delta\rangle^2\right]^{1/2}&\lsim&0.003.
\eeqa

In models of alignment,
\beqa
|K_{21}^{u_L}K_{11}^{u_L*}|&\approx&|V_{us}|\sim0.23,\no\\
|K_{21}^{u_R}K_{11}^{u_R*}|&\sim&(m_u/m_c)/|V_{us}|\sim0.009,
\eeqa
where we used $m_u/m_c\approx0.002$. Comparing to Eq. (\ref{upabsd}),
we find the following upper bounds on mass splittings:
\beqa
\frac{\Delta\tilde m^2_{Q_2Q_1}}{\tilde m^2_Q}&\leq&0.56,\no\\
\left[\frac{\Delta\tilde m^2_{Q_2Q_1}}{\tilde m^2_Q}
\frac{\Delta\tilde m^2_{U_2U_1}}{\tilde m^2_U}\right]^{1/2}&\lsim&0.17.
\eeqa
Furthermore, in models of alignment, the phases are assumed to be of
order one. Taking maximal phases, we obtain from Eq. (\ref{upimad})
\beqa
\frac{\Delta\tilde m^2_{Q_2Q_1}}{\tilde m^2_Q}&\leq&0.23,\no\\
\left[\frac{\Delta\tilde m^2_{Q_2Q_1}}{\tilde m^2_Q}
\frac{\Delta\tilde m^2_{U_2U_1}}{\tilde m^2_U}\right]^{1/2}&\lsim&0.071.
\eeqa

Taking \cite{Raz:2002zx} $\tilde m_Q=\frac12(\tilde m_{Q_1}+\tilde
m_{Q_2})$ and similarly for the SU(2)-singlet squarks, we find that
we thus have an upper bound on the splitting between
the first two squark generations:
\beqa\label{conali}
\frac{m_{\tilde Q_2}-m_{\tilde Q_1}}{m_{\tilde Q_2}+m_{\tilde Q_1}}
&\lsim&0.05-0.14,\no\\
\frac{m_{\tilde u_2}-m_{\tilde u_1}}{m_{\tilde u_2}+m_{\tilde u_1}}
&\lsim&0.02-0.04.
\eeqa
The first bound applies to the up squark doublets, while the second to
the average of the doublet mass splitting and the singlet mass
splitting. The range in each of the bounds corresponds to values of
the phase between zero and maximal. We can thus make the following
conclusions concerning models of alignment:
\begin{enumerate}
  \item The mass splitting between the first two squark doublet
    generations should be below 14\%. For phases of order one, the
    bound is about $2-3$ times stronger.
    \item In the simplest models of alignment, the mass splitting
      between the first two squark generations should be smaller than
      about four percent.
      \item The second (stronger) bound can be avoided in
          more complicated models of alignment, where
          holomorphic zeros suppress the mixing in the
          singlet sector. \item While renormalization group
          evolution (RGE) effects can
provide some level of universality, even for anarchical
boundary conditions, the upper bound (\ref{conali})
requires not only a high scale of mediation
\cite{Nir:2002ah} but also that, at the scale of mediation,
the gluino mass is considerably higher than the squark
masses.
\end{enumerate}

In any model where the splitting between the first two squark
doublet generations is larger than ${\cal O}(y_c^2)$,
$|K_{21}^{u_L}-K_{21}^{d_L}|=\sin\theta_c=0.23$. Given the
constraints from $\Delta m_K$ and $\epsilon_K$ on
$|K_{12}^{d_L}|$, one arrives at a constraint very similar to
the first bound in Eq. (\ref{conali}). We conclude that the
constraints on the level of degeneracy between the squark
doublets (stronger than five to fourteen percent) apply to any
supersymmetric model where the mass of the first two squark
doublet generations is below TeV. It is suggestive that the
mechanism that mediates supersymmetry breaking is
flavor-universal, as in gauge mediation.

\section{Warped extra dimensions}
\label{sec:rs}
Randall-Sundrum (RS) models of warped extra dimensions predict
that there are new tree level contributions to
$D^0-\overline{D^0}$ mixing from the exchange of Kaluza-Klein
(KK) gluons~\cite{Agashe:2004cp}. The flavor suppression of
such contributions is determined, up to ${\cal O}(1)$
uncertainties from the five-dimensional Yukawa couplings, by
the values of the quark wave-functions on the IR brane,
$f_{Q_i}$ ($f_{u_i,d_i}$), for the left-handed (right-handed)
fields. These wave-functions, which depend on the fermion bulk
masses $c_i$, can be estimated from the quark flavor
parameters,
\beq
|V_{ij}|\sim f_{Q_i}/f_{Q_j},\ \ \ y_{u_i}\sim f_{Q_i}f_{u_i},\ \ \
y_{d_i}\sim f_{Q_i}f_{d_i},
\eeq
where $y_{q_i}$ are the four-dimensional Yukawa couplings.

A question that is crucial in many respects and, in particular,
for the discovery potential of the LHC, is that of the lower
bound on the mass of the KK excitations. There are bounds from
electroweak precision measurements on the KK mass scale of
order 3~TeV. The weakest flavor constraints arise in models
where there is an alignment of the down sector flavor
parameters such that contributions to flavor changing neutral
currents in the down sector are highly
suppressed~\cite{Align,Csaki:2008eh,kkgluon2}. In that case,
the bounds from $D^0-\overline{D^0}$ mixing play a crucial
role, since the up sector tends to possess an anarchical
structure as in the generic models~\cite{Agashe:2004cp}.

We first consider the case where the Higgs is localized on the
IR brane. The KK gluon exchange contributes to the various
terms of Eq.~(\ref{nonren}) by mediating four-quark
interactions:
\beqa
\Lambda_{\rm NP}&=&M_G,\no\\
z_1&\sim&\frac{g_{s*}^2}{6} \gamma(c_{Q_2})^2
(V_{ub}V^*_{cb})^2f_{Q_3}^4,\no\\
z_4&\sim&\frac{g_{s*}^2}{Y_*^2} \gamma(c_{Q_2}) \gamma(c_{u_2})
y_u y_c,
\eeqa
where we listed the two operators that yield the strongest
constraints. Here $M_G$ is the mass of the KK gluon, $g_{s*}$
is the bulk SU(3) gauge coupling, $Y_*$ is the typical size of
the (presumably anarchical) entries in the IR brane localized
Yukawa interactions and $\gamma(c)$ is a correction function to
the overlap of the quarks with the first KK gluon, given
by~\cite{kkgluon1,kkgluon2}
\beq
\gamma(c)=\frac{\sqrt{2}}{J_1(x_1)} \frac{0.7}{6-4c} \left(
1+e^{c/2} \right) \, ,
\eeq
with $x_1 \approx 2.4$ being the first root of the Bessel function
$J_0(x_1)=0$.

For the purpose of a quantitative analysis, we take $g_{s*}=3$,
obtained by matching to the 4D coupling at
one-loop~\cite{Agashe:2008uz}. We also use $c_{Q_2}=0.58$ and
$c_{u_2}=0.53$, which are reasonable representative values for
the relevant range of the parameters $f_{Q_3}$ and $Y_*$ (in
which the result can change by only a few percent). The value
of $f_{Q_3}=0.4\,(\sqrt{2})$ represents a profile that is
fairly flat, $c_{Q_3}=0.42$ (sharply localized in IR,
$c_{Q_3}=-0.5$)~\footnote{We thank Kaustubh Agashe for pointing
out an inaccuracy in these values in the previous version of
the paper.}. In any case, our strongest constraint, which comes
from $z_4$, is independent of $f_{Q_3}$. For the quark sector
parameters, we use
\beq\label{rsquark}
y_u=6.1\times10^{-6},\ \ y_c=2.95\times10^{-3},\ \
|V_{ub}V^*_{cb}|=1.6\times10^{-4},
\eeq
where we evaluate $m_{u,c}$ at 3~TeV with SM-RGE, given in
Ref.~\cite{Xing:2007fb} (the experimental bounds are also
calculated at 3~TeV), and use the central values of the CKM
elements given in Ref.~\cite{Amsler:2008zzb}. We present our
results for the lower bound on the KK scale (that is, the mass of
the KK gluon) in Table~\ref{tab:rszi}. This analysis can be used
to obtain a lower bound on $Y_*$, assuming $M_G=3$ TeV, which is
the minimal value allowed by electroweak precision constraints.
This is also shown in Table~\ref{tab:rszi}.
\begin{table}[t]
\caption{The $z_i$ parameters in a generic RS model. The bound on
$Y_*$ is for $M_G=3$ TeV. Both bounds correspond to maximal phase,
and are relaxed by a factor $\sim2.4$ for a vanishing phase.}
\label{tab:rszi}
\begin{center}
\begin{tabular}{cc|c|c} \hline\hline
\rule{0pt}{1.2em}%
Parameter & Numerical estimate &  $(M_G)_{\rm min}[\mathrm{TeV}]$
& $(Y_*)_{\rm min}$ \cr \hline $z_1$ & $\
5.8\times10^{-7}f_{Q_3}^4\ $ & $0.73 f_{Q_3}^2$ & - \cr $z_4$ & $\
{2.2\times10^{-7}/ Y_*^2}\ $ & ${4.9/ Y_*}$ & $1.6$ \cr \hline\hline
\end{tabular}
\end{center}
\end{table}

Note that in Ref.~\cite{Blum:2009sk}, a much stronger bound was
derived based on the same experimental data. This difference is the
result of a combination of two $\mathcal{O}(1)$ factors: First,
$g_{s*}$ was taken to be 6 and not 3 (based on tree-level instead of
one-loop matching), and second the CKM factor was roughly evaluated by
$\lambda_C^5$, instead of $|V_{ub}V^*_{cb}|$ in the current paper (the
latter is more appropriate in the context of constraining models with
alignment in the down sector). These differences reflect the fact that
$\mathcal{O}(1)$ uncertainties are always present in the RS framework,
as a result of its limited predictive power.

The bounds in Table~\ref{tab:rszi} can be further relaxed by
considering a bulk Higgs (note that the dimension of the Yukawa
coupling changes in this case, so that $Y_*$ is given in units of
$\sqrt{k}$). The couplings of the light quarks with a bulk
Higgs are \emph{enhanced}. The RS contribution to $z_1$ does not
change, to leading order, since the effect of a bulk Higgs on the
rotation angles $f_i/f_j$ is subleading. However, the mass relations
of the form $y_i \approx Y_* f_{Q_i} f_{u_i}$, used to obtain the
contribution to $z_4$, are altered in this case.  This change can be
expressed by a function that corrects for the overlap of the
wavefunctions of the Higgs with two zero-mode quarks, relative to the
IR Higgs case~\cite{Gedalia:2009ws}:
\beq
y_i \approx Y_* f_{Q_i} f_{u_i} r^H_{00}(\beta, c_{Q_i},
c_{u_i}) \, , \qquad r^H_{00}(\beta, c_{Q_i}, c_{u_i})=
\frac{\sqrt{2(1+\beta)}}{2+\beta-c_{Q_i}-c_{u_i}} \, ,
\eeq
where $r^H_{00}$ depends on the Higgs profile in the bulk,
parameterized by $\beta=\sqrt{4+\mu^2}$ ($\mu$ is the bulk mass
of Higgs in units of $k$). Hence, $z_4$ is given in this case
by
\beq
z_4 \sim \frac{g_{s*}^2}{Y_*^2} \frac{\gamma(c_{Q_2})
\gamma(c_{u_2})}{r^H_{00}(\beta, c_{Q_1}, c_{u_1})
r^H_{00}(\beta, c_{Q_2}, c_{u_2})} y_u y_c \, .
\eeq
For a Higgs maximally spread into the bulk (that is, saturating
the anti-de Sitter stability bound~-- $\beta=0$), the bound on
$M_G$ is reduced by a factor of $\sim 2$, to $2.4/Y_*$ TeV.

We learn that the recent measurements of $D^0-\overline{D^0}$
mixing impose additional constraints on the RS model. In
particular, given an IR Higgs, a 3 TeV KK scale requires $Y_*
\gtrsim 1.6$, which is close to the perturbativity bound $Y_*
\lesssim 2\pi/N_{\rm KK}$, where $N_{\rm KK}$ stands for the
number of KK state below the theory's UV cutoff. That has also
implications for alignment models~\cite{Align}, where the
larger the value of $Y_*$ is, the larger the next to leading
order corrections are, which spoils the alignment.

\section{Conclusions}
\label{sec:con}
Recent bounds on CP violation in $D^0-\overline{D^0}$ mixing are
particularly significant, because -- unlike the mass splitting and
the width splitting -- there is no standard model contribution that
can interfere with the new physics. We studied the implications of
these measurements to various frameworks of new physics.

For generic models, we obtained the following results:
\begin{itemize}
\item Generic new physics that contributes to the operators
(\ref{nonren}) at tree level with couplings of ${\cal O}(1)$
must lie at a very high scale, $\Lambda_{\rm NP}\gsim
(4-10)\times10^3\ TeV$.
\item Generic new physics that contributes to the operators
(\ref{nonren}) at the loop level with effective coupling of
${\cal O}(\alpha_2^2)$ (similar to the SM)
must lie at a high scale, $\Lambda_{\rm NP}\gsim
(1-3)\times10^2\ TeV$.
\item New physics at or below the TeV scale must have a highly
  suppressed coupling, {\it e.g.}, $z_1\lsim(1-6)\times10^{-7}$
  and $z_4\lsim(1-6)\times10^{-8}$.
If the new physics contribution is loop-suppressed by
${\cal O}(\alpha_2^2)$, then the flavor suppression has to be
as strong as $10^{-4}-10^{-5}$.
\item Neither electroweak loop suppression nor alignment of order
  $\sin\theta_c$ are sufficient to allow new physics at the TeV
  scale. There must be some level of degeneracy -- stronger than
  ${\cal O}(0.1)$ -- to allow that.
\end{itemize}

For models with minimal flavor violation (MFV), we reached the
following conclusions:
\begin{itemize}
\item MFV models with two Higgs doublets can contribute to
  $D^0-\overline{D^0}$ mixing up to ${\cal O}(0.1)$ of the
  experimental value for very large $\tan\beta$.
\item Single Higgs doublet models, and two Higgs doublet models with
  small $\tan\beta$, contribute at ${\cal O}(10^{-7})$.
\item The CP violating part of these contributions is not suppressed
  compared to the CP conserving part.
\end{itemize}
Our findings imply that MFV models with very large $\tan\beta$ will be
probed  once the experimental sensitivity to CP violation in mixing
reaches the ten percent level.

For supersymmetric models with quark-squark alignment, we learn the
following:
\begin{itemize}
    \item The mass splitting between the first two squark doublet
    generations should be below 14\%. For phases of order one, the
    bound is two to three times stronger.
    \item In the simplest models of alignment, the mass splitting
      between the first two squark generations should be smaller than
      about four percent.
      \item The second (stronger) bound can be avoided in more
      complicated models of alignment where holomorphic zeros suppress
      the mixing in the singlet sector.
      \item
While RGE effects can provide some level of universality, even for
anarchical boundary conditions, the upper bound (\ref{conali}) requires
not only a high scale of mediation \cite{Nir:2002ah} but also that, at the
scale of mediation, the gluino mass is considerably higher than the
squark masses \cite{Blum:2009sk}.
\end{itemize}
For models of warped extra dimensions where alignment is used to relax
the bounds from $K^0-\overline{K^0}$ mixing, we find the following:
\begin{itemize}
  \item The lower bound on the KK-gluon mass is pushed to
      be above 2.5 (10) TeV for the maximally (minimally)
      composite top case depending on the size of the 5D
      Yukawa coupling assuming at least three KK states.
      The flavor constraints are then stronger than those
      from the electroweak precision measurements for a
      large portion of the parameter space.
\end{itemize}

\section*{Acknowledgements}
This work is supported by the United States-Israel Binational Science
Foundation (BSF), Jerusalem, Israel.  The work of YG is supported by
the NSF grant PHY-0757868. The work of YN is supported by the Israel
Science Foundation (ISF) under grant No.~377/07, the German-Israeli
foundation for scientific research and development (GIF), and the
Minerva Foundation. The work of GP is supported by the Peter and
Patricia Gruber Award.


\end{document}